\documentclass{aastex61}

\usepackage{amsmath,amssymb,amsfonts}
\usepackage{color}

\usepackage[T1]{fontenc}
\usepackage{txfonts}
\usepackage[]{graphicx}
\def\comment#1{}

\submitjournal{ApJ}

\shorttitle{Dark matter: An efficient catalyst for IMRIs}
\shortauthors{Yue, Han \& Chen, 2018}

\begin{document}

\title{Dark matter: An efficient catalyst for intermediate-mass-ratio-inspiral events}

\correspondingauthor{Wen-Biao Han}
\email{wbhan@shao.ac.cn}

\author{Xiao-Jun Yue}
\affiliation{Shanghai Astronomical Observatory, Shanghai, China, 200030}
\affiliation{College of Information and Computing, Taiyuan University of Technology, Taiyuan 030024, China}

\author{Wen-Biao Han}
\affiliation{Shanghai Astronomical Observatory, Shanghai 200030,People's Republic of China, }
\affiliation{School of Astronomy and Space Science, University of Chinese Academy of Sciences, Beijing 100049, China}

\author{Xian Chen}
\affiliation{Astronomy Department, School of Physics, Peking University,
Beijing 100871, China}
\affiliation{Kavli Institute for Astronomy and Astrophysics at
Peking University, Beijing 100871, China}




\begin{abstract}
Gravitational waves (GWs) can be produced if a stellar compact object, such as
a black hole (BH) or neutron star, inspirals into an intermediate-massive
black hole (IMBH) of $(10^3 \sim 10^5)\,M_\odot$. Such a system may be produced
in the center of a globular cluster (GC) or a nuclear star cluster (NSC), and
is known as an intermediate- or extreme-mass-ratio inspiral (IMRI or EMRI).
Motivated by the recent suggestions that dark matter minispikes could form
around IMBHs, we study the effect of dynamical friction against DM on the
merger rate of IMRIs/EMRIs. We find that the merger timescale of IMBHs with BHs
and NSs would be shortened by two to three orders of magnitude.  As a result,
the event rate of IMRIs/EMRIs are enhanced by orders of magnitude relative to
that in the case of no DM minispikes. In the most extreme case where IMBHs are
small and the DM minispikes have a steep density profile, all the BH in GCs and
NSCs might be exhausted so that the mergers with NSs would dominate the current
IMRIs/EMRIs.  Our results suggest that the mass function of the IMBHs below
$10^4 \,M_\odot$ would bear imprints of the distribution of DM minispikes
because these low-mass IMBHs can grow efficiently in the presence of DM
minispikes by merging with BHs and NSs. Future space-based GW detectors, like
LISA, Taiji, and Tianqin, can measure the IMRI/EMRI rate and hence constrain
the distribution of DM around IMBHs.
\end{abstract}




\section{Introduction} \label{S1}

Cosmological and astrophysical observations have provided reliable evidence for
the existence of dark matter (DM). It is important to understand the
distribution of DM in different astrophysical systems.  \citet{I1} first
pointed out a universal density profile for DM halos called the NFW profile.
\citet{I2} suggested that the adiabatic growth of a supermassive black hole
(SMBHs) with a mass $M\sim10^{6-9}M_\odot$  at the center of a galaxy can
create a high density cusp of DM called the DM "spike." However, the DM spikes
could be removed by astrophysical processes such as galaxy mergers, formation
of off-centered seed BHs, and scattering with the surrounding stars
\citep{I3,I4,I5,I6}.

On the other hand, DM minispikes may exist around an intermediate-massive black
hole (IMBH) with a mass of $M\sim 10^{3-5}M_\odot$. In particular, it has been
shown that spinning IMBHs could have formed minispikes more easily
\citep{Ferrer17}.  Although the formation process of IMBHs is still unclear,
there is an increasing number of evidence suggesting that they should exist in
the centers of globular clusters (GCs) or the nuclear stellar clusters (NSCs)
of dwarf galaxies.  The active galactic galactic nuclei (AGNs) in dwarf
galaxies provide robust evidence supporting the existence of $\approx
10^5M_\odot$ BHs. Another strong piece of evidence is the ultra-luminous X-ray sources
\citep{ Farell et al.09,2010;Feng&Soria 2011}. The final proof of the existence of IMBHs
requires detection of the innermost stellar kinematics in GCs or NSCs, which is
still difficult to achieve today. Nevertheless, if minispikes exist around
IMBHs, they are less likely to be destroyed by galaxy mergers because their
hosts may not have experienced major mergers in the past \citep{I7,I8}.

The recent detection of gravitational waves (GWs) opened a new possibility of
proving the existence of IMBHs.  If a stellar-mass black hole (BH) orbits
around an IMBH or SMBH, the two objects form an intermediate-mass-ratio
inspiral (IMRI, of a mass ratio of $10^{-2} \sim 10^{-3}$) or an extreme-mass-ratio inspiral (EMRI, with a mass ratio of $10^{-4} \sim 10^{-6}$). Such a
system is an ideal source for the space-borne detectors such as LISA
\citep{LISAa,LISAb}, Tianqin \citep{taiji},  and Taiji \citep{tianqin}. It was
estimated in \citet{I11} that there are 10 IMRIs in the LISA band at any time.
Moreover, if the mass of the IMBH is $\lesssim 10^3 M_\odot$, the final merger
phase of the IMBH and the stellar BH could be detected by the advanced LIGO
\citep{I13,I12}.

Detecting IMBHs via GWs would also allow us to test the existence of DM
minispikes.  For example, \citet{I14} showed that the additional gravitational
attraction due to the DM minispike can affect the waveform of an EMRI, and the
deviate of the observed waveform from a standard EMRI template is detectable by
LISA. Furthermore, \citet{I15} showed that the dynamical friction of the DM
could also induce an observable effect on the EMRI waveform. In addition,
\citet{I16} considered the combined effect of gravitational pulling, dynamical
friction, and the accretion of DM, and showed that the dynamical friction is
predominant.

Based on the results of the previous studies, here we calculate the merger
times for the IMRIs/EMRIs embedded in DM minispikes (Sec.\ref{S2}). We use them
to further derive the event rates in GCs (Sec.\ref{s4}) and the NSCs in dwarf
galaxies (Sec.\ref{s5}). We discuss the robustness of our results in
Section~\ref{s6} and provide our conclusions in Section~\ref{s7}.

\section{Merger time in DM minispikes} \label{S2}

Our DM minispike is the same as that derived in \citet{I15}, whose density
$\rho$ follows a power-law distribution of
\begin{equation}
\rho(r)=\rho_{\rm
sp}\left(\frac{r_{\rm sp}}{r}\right)^\alpha,\label{1}
\end{equation}
where $r_{\rm sp}$ is a typical scale radius normally related to the influence
radius $r_h$ of the central massive black hole (MBH) as $r_{\rm sp}\sim 0.2
r_{h}$ and $\rho_{\rm sp}$ is the DM density at $r=r_{\rm sp}$.  The influence
radius can be calculated from the equation $4\pi\int_0^{r_h}\rho_{\rm DM}r^2
dr=2M_{\rm BH}$ where $M_{\rm BH}$ denotes the mass of the central MBH.
Following the derivation in \citet{I15}, we use $r_{\rm sp} \simeq 0.54\, {\rm
pc}$ and $\rho_{\rm sp}=226\,M_\odot/{\rm pc}^3$ throughout this paper.  We
note that in principle $\rho_{\rm sp}$ and $r_{\rm sp}$ should depend on the
mass of the central MBH, but the relationship has not been derived in the
literature . Therefore, for simplicity, here we use the same values of
$\rho_{\rm sp}$ and $r_{\rm sp}$ for the IMBHs in the mass range of
$10^{3-5}M_\odot$.

As for the power-law index $\alpha$, it can be derived according to the model
of adiabatic growth of massive black holes \citep[MBHs, see][]{Young}.  If the
initial DM halo, prior to the formation of the MBH, has a NFW profile with an
initial power-law index of $\alpha_{\rm ini}=1$ \citep{I1}, the index after the
adiabatic growth of the MBH is $\alpha=(9-2\alpha_{\rm ini})/(4-\alpha_{\rm
ini})=7/3$ \citep{I4,I17}.  Alternatively, if the initial halo has a uniform
density distribution, the final power-law index is  $\alpha=1.5$
\citep{I4,I17}. For these reasons, we will assume $1.5<\alpha<7/3$ in the
following analysis.

We now consider a binary composed of an IMBH with a mass of $M_{\rm BH}\sim
10^{3-5} M_\odot$ and a compact object such as a stellar-mass BH with a mass of
$\mu=10M_\odot$ or a neutron star (NS) with $\mu=1.4 M_\odot$.  The binary is
embedded in the center of the DM minispike.  Since the mass of the secondary
compact object is much smaller than the central IMBH, the reduced mass of the
binary is approximately $\mu$ and the center-of-mass is approximately at the
position of the IMBH.

In the Newtonian formalism, the motion of the secondary compact object can be
decomposed into a radial component and a tangential one.  The equation of
motion in the radial direction is $\mu\ddot{r}-\mu r\dot{\theta}^2=F$, where
$F$ is the gravitational force imposed by the IMBH as well as the DM minispike.
Inside the radius of the innermost stable orbit, i.e., $r<r_{\rm ISCO}$, there
is no stable orbit for DM particles so that they all fall into the central
hole.  As a result, the DM has a hollow distribution around the IMBH and this
needs to be accounted for in the calculation of the gravitational force. In the
end the equation in the radial direction can be written in the form \citep{I14}
\begin{equation}
\mu\ddot{r}-\mu r\dot{\theta}^2=-\frac{G\mu M_{\rm eff}}{r^2}-\frac{\mu F}{r^{\alpha-1}},\label{2}
\end{equation}
where
\begin{eqnarray}
M_{\rm eff}&&=
\begin{cases}
M_{\rm BH}-M_{\rm DM}(<r_{\rm min}),& \text{$r_{\rm min}<r<r_{\rm sp}$},\\
M_{\rm BH},& \text{$r<r_{\rm min}$},\\
\end{cases}\\\label{3}
F&&=
\begin{cases}
G r_{\rm min}^{\alpha-3}M_{\rm DM}(<r_{\rm min}),& \text{$r_{\rm min}<r<r_{\rm sp}$},\\
0,& \text{$r<r_{\rm min}$}.
\end{cases}\label{4}
\end{eqnarray}
In the above equations, $r_{\rm min}=r_{\rm ISCO}=6GM_{\rm BH}/c^2$
 is the radius of the innermost stable circular orbit
(ISCO) and $M_{\rm DM}(<r_{\rm min})=4\pi r_{\rm sp}^\alpha\rho_{\rm sp}r_{\rm
min}^{\alpha-3}/(3-\alpha)$ is the DM contained in $r_{\rm ISCO}$. The first
term on the right-hand side of Equation~(\ref{2}) is the effective mass of IMBH
corrected by DM. The second is the gravitational effect of DM. For a circular
orbit, $\ddot{r}=0$, and
\begin{equation}
\dot{\theta}=\omega_s=\sqrt{\frac{GM_{\rm
eff}}{r^3}+\frac{F}{r^\alpha}}\label{40}
\end{equation}
is the orbital frequency.

As the secondary compact object moves in the DM minispike, its orbital energy
is lost due to GW radiation and the dynamical friction against the DM background.
The equation of energy balance is
\begin{equation}
-\frac{dE_{\rm orbit}}{dt}=\frac{dE_{\rm GW}}{dt}+\frac{dE_{\rm DF}}{dt},\label{5}
\end{equation}
where the terms are explained in as follows. The orbit energy is
\begin{eqnarray}
E_{\rm orbit}&&=\frac{1}{2}\mu v^2-\frac{G\mu M_{\rm eff}}{r}+\frac{1}{2-\alpha}\frac{\mu F}{r^{\alpha-2}}\nonumber\\
&&=-\frac{G\mu M_{\rm eff}}{2r}+\frac{4-\alpha}{2(2-\alpha)}\frac{\mu F}{r^{\alpha-2}},\label{6}
\end{eqnarray}
where $v$ is the circular velocity of the small body. The rate of energy loss via
GWs is
\begin{equation}
\frac{dE_{\rm DF}}{dt}=\frac{32}{5}\frac{G\mu^2}{c^5}r^4\omega_s^6.\label{7}
\end{equation}
The force due to dynamical friction is $f_{\rm DF}=4\pi G^2\mu^2\rho_{\rm
DM}(r)\ln\Lambda/v^2$ and the corresponding energy dissipation rate is
\begin{equation}
\frac{dE_{\rm DF}}{dt}=vf_{\rm DF}=4\pi G^2\frac{\mu^2\rho_{\rm DM}(r)}{v}\ln\Lambda \,. \label{8}
\end{equation}
where $\ln\Lambda$ is the Coulumb logarithm and we assume $\ln\Lambda=10$.
Replacing the terms in  Equation~(\ref{5}) using Equations~(\ref{1}),
(\ref{6}), (\ref{7}), and (\ref{8}), we find
\begin{eqnarray}
\frac{dr}{dt}=&&-\left(\frac{GM_{\rm eff}}{2r^2}+\frac{4-\alpha}{2}\frac{F}{r^{\alpha-1}}\right)^{-1}\nonumber\\
&&\times\left[\frac{32}{5}\frac{G\mu}{c^5}r^4\omega_s^6+\frac{4\pi G^2\mu\rho_{\rm sp}r_{\rm sp}^\alpha \ln\Lambda}{r^{\alpha+1}\omega_s}\right],\label{9}
\end{eqnarray}
where $\omega_s$ is derived in Equation~(\ref{40}). For example, if the DM
minispike is absent, $F=0$ and the second term in the brackets vanishes, and
hence Equation~(\ref{9}) reduces to
\begin{equation}
\frac{dr}{dt}=-\frac{64}{5}\frac{G^3\mu M^2}{c^5 r^3},\label{10}
\end{equation}
which is the well-known Peters formula \citep{Peters63}.

\begin{figure}[!h]
\begin{center}
\includegraphics[height=4.0in]{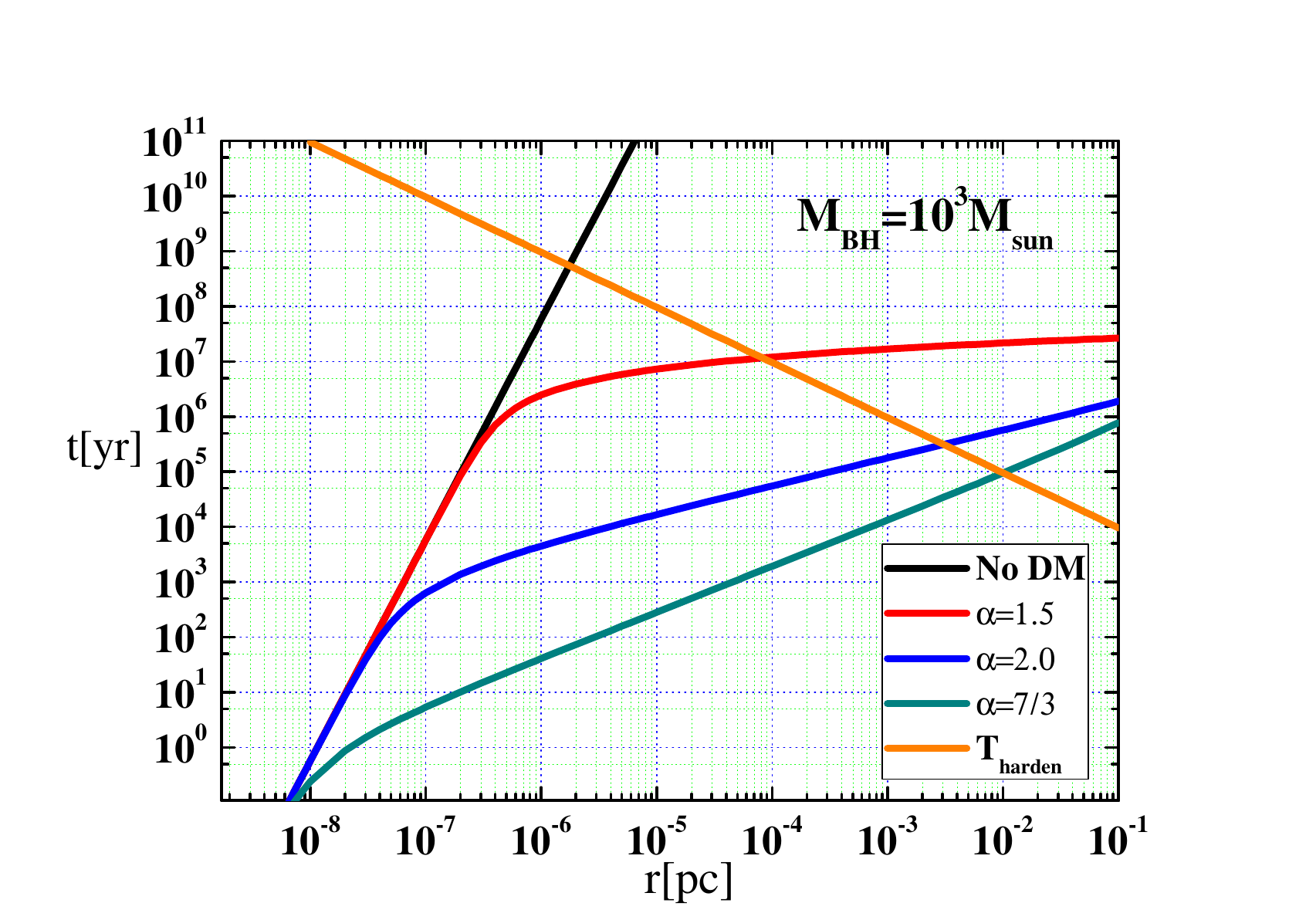}
\caption{Merger time for an IMRI with $\mu=10\,M_\odot$ and $M_{\rm BH}=10^3
M_\odot$ computed using different DM density profiles (the black, red, blue,
and green curves), as well as the hardening time due to scattering away DM
particles (orange line).  }  \label{fig01}
\end{center}
\end{figure}

\begin{figure}[!h]
\begin{center}
\includegraphics[height=4.0in]{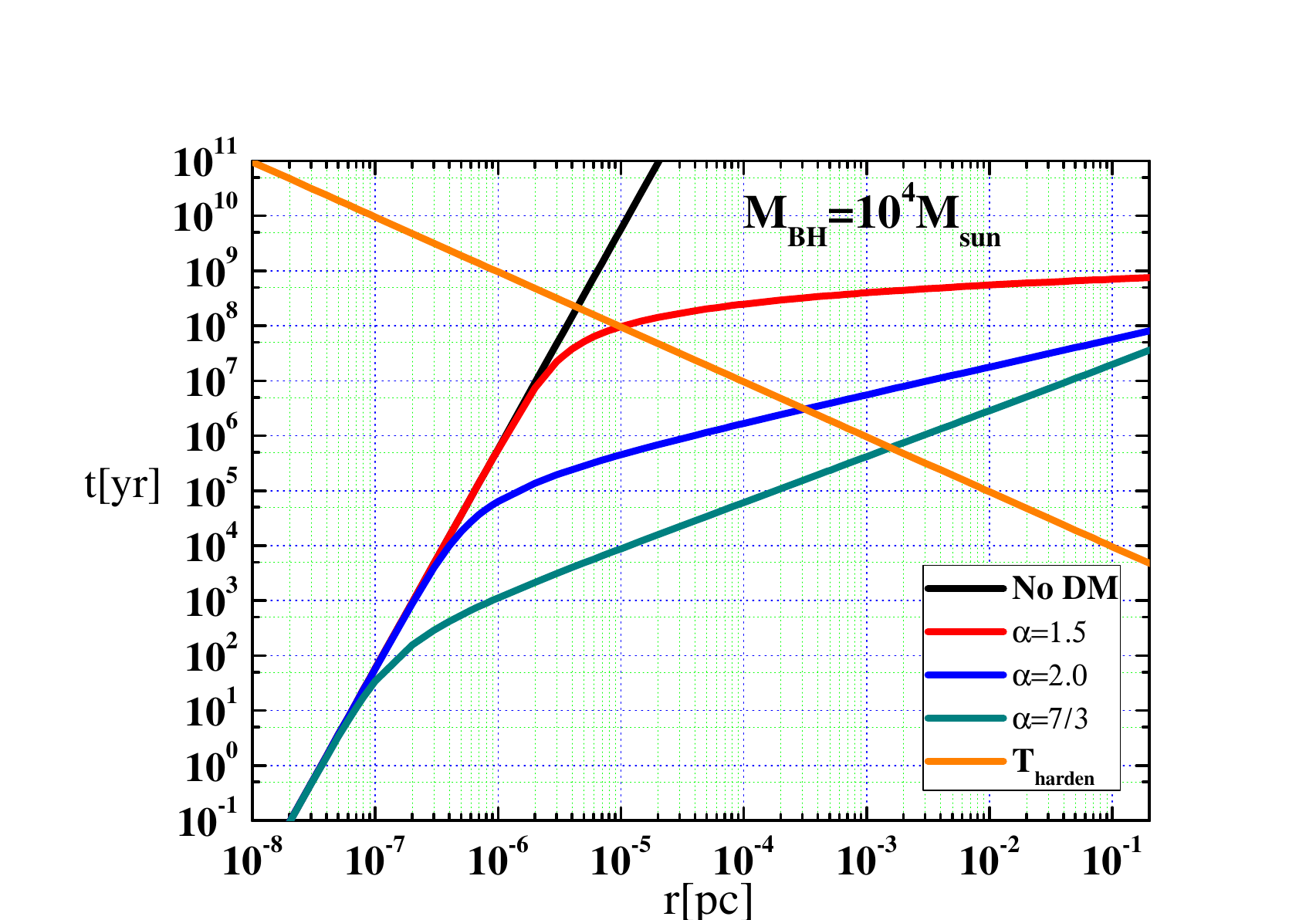}
\caption{Same as Figure~\ref{fig01} but for $M_{\rm BH}=10^4\,M_\odot$.}
\label{fig02}
\end{center}
\end{figure}

\begin{figure}[!h]
\begin{center}
\includegraphics[height=4.0in]{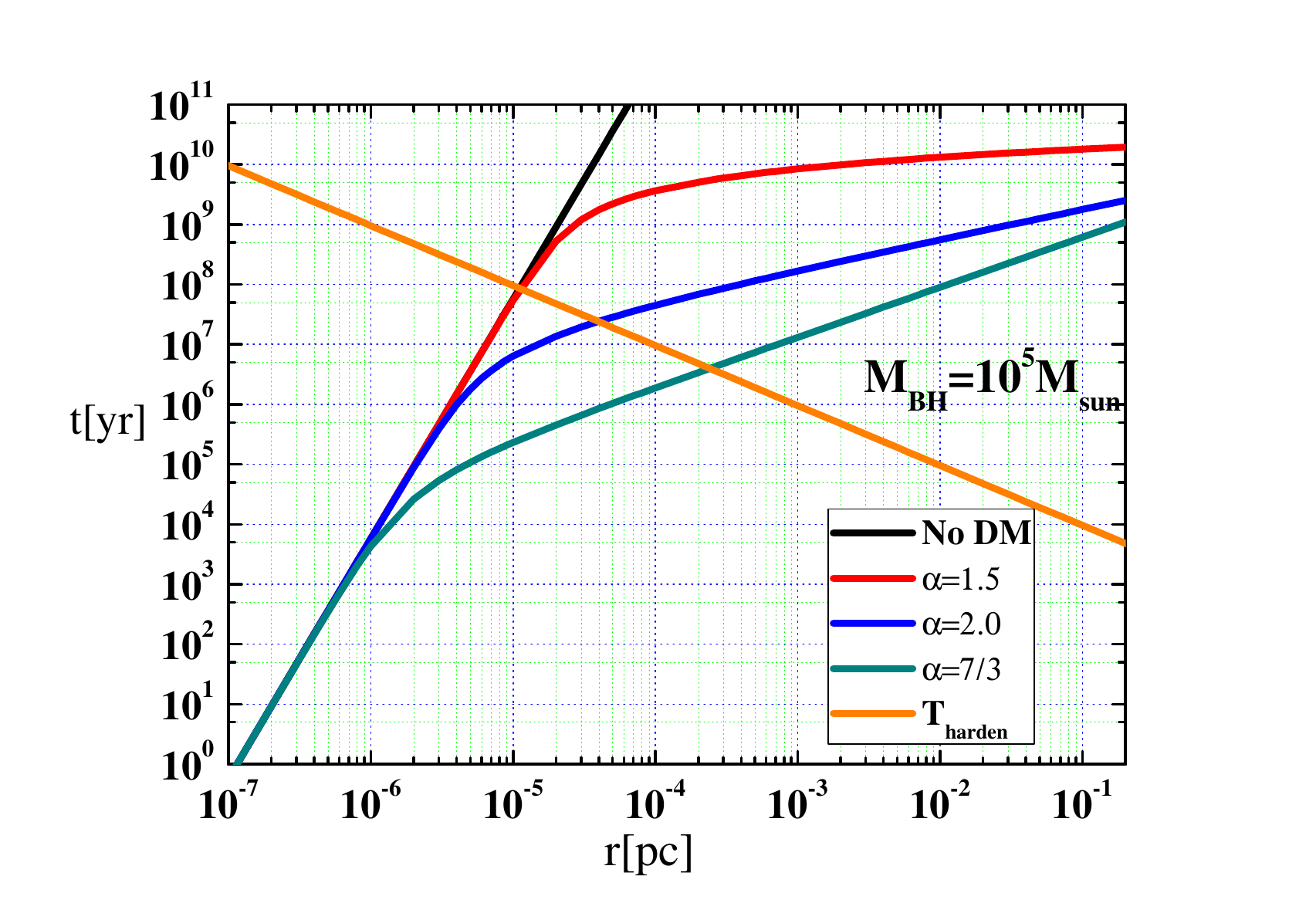}
\caption{Same as Figure~\ref{fig01} but for $M_{\rm BH}=10^5\,M_\odot$.}
\label{fig03}
\end{center}
\end{figure}

Using Equations~(\ref{9}) or(\ref{10}), we can derive the merger time from an
initial radius of $r$ to the final radius $r_{\rm ISCO}$, which is
\begin{equation}
T_{\rm merge}=\int_{r}^{r_{\rm ISCO}}\frac{dt}{dr}dr.\label{11}
\end{equation}
Figures~\ref{fig01}-\ref{fig03} illustrate our results of the merger times for
different parameters $\alpha$ and different masses of IMBH.  It is clear that
the presence of DM minispikes greatly reduces the merger times of IMRIs.
We note that $T_{\rm merge}$ is approximately the evolution timescale at
$r$ due to GW radiation and DM friction, because the time spent at $r$ is much
longer than that at $r_{\rm ISCO}$.

Besides the dynamical friction against the DM background, another competing
mechanism that could also extract the orbital energy from the binary of IMBH
and stellar BH is ``dynamical hardening'' \citep{heggie75,I18,I13}.  According to
this mechanism, a stellar interloper could interact with the binary in a
complicated way such that by the end of the interaction the interloper is
reejected into the background, taking away a fraction of the energy and
angular momentum from the binary.

To study the relative efficiency of the dynamical hardening, we adopt the method
of \cite{I19} and calculate the hardening rate using
\begin{equation}
\frac{da}{dt}=-\frac{G H\rho a^2}{\sigma}\label{12},
\end{equation}
where $a$ is the semi-major axis of the binary, $H\sim 15$ is a constant,
$\rho$ is the typical density of the background stars, and $\sigma$ is the
velocity dispersion of these stars.  For simplicity, we assume that $\rho$ is
constant, which mimics the core profile observed in the centers of many massive
GCs.  Moreover, we do not consider a significant eccentricity $e$ for our
binaries, i.e., we have $a=r$. Finally, the hardening timescale from the
initial radius $r_i$ to the final one $r$ is
\begin{equation}
T_{\textrm{harden}}=\int^{r}_{r_i}\left(\frac{dt}{dr}\right)_{\textrm{harden}}dr\label{13},
\end{equation}
where $(dt/dr)_{\textrm{harden}}$ is from Equation~(\ref{12}).  We note that
the choice of $r_i$ is not important for the calculation of
$T_{\textrm{harden}}$ because $(dt/dr)_{\textrm{harden}}$ quickly approaches
$0$ as $r$ increases. For this reason, we assume $r_i=\infty$ in our
calculations and  the value given by Equation~(\ref{12}) is approximately
the hardening timescale at the radius $r$, since the IMRIs/EMRIs evolve much
slower at smaller $r$. By solving for the radius where
$T_{\textrm{harden}}=T_{\rm merge}$, we can find the transition radius of the
two processes. These critical radii are shown in
Figures~\ref{fig01}-\ref{fig03} as the intersections of the lines of $T_{\rm
merge}$ and $T_{\rm harden}$.  

For example, if we use the typical parameters for GCs, i.e., $\sigma=10\,{\rm
km\,s^{-1}}$, $\rho=n m_\ast$ where $n=10^{5.5}\,{\rm pc}^{-3}$ is the number
density of stars, and $m_\ast=0.5\,M_\odot$ is the average mass for a single
star \citep{I20}, we can compute the corresponding $T_{\textrm{harden}}$ ,which
are shown in Figures~(\ref{fig01})-(\ref{fig03}) as the
oranges lines.   Comparing these results with $T_{\rm merge}$, we find that in
general the effect of dynamical hardening is more important at large radius and
subsides as the semi-major axis of the binary shrinks.

Since both the DM and dynamical hardening are affecting the evolution, the
total shrinking rate due to the two effects combined is
\begin{equation}
\left(\frac{dr}{dt}\right)_{\textrm{total}}=\left(\frac{dr}{dt}\right)_1+\left(\frac{dr}{dt}\right)_2,\label{14}
\end{equation}
where $(dr/dt)_1$ is the rate due to GW and DM, which is computed from
Equation~(\ref{9}), and $(dr/dt)_2$ is the hardening rate from
Equation~(\ref{12}). Correspondingly, the total merger time combining the two
effects is
\begin{equation}
T=\int_{r_i}^{r_{\textrm{ISCO}}}\left(\frac{dt}{dr}\right)_{\textrm{total}}dr=\int_{r_i}^{r_{\textrm{ISCO}}}\frac{1}{\left({dr}/{dt}\right)_1+\left({dr}/{dt}\right)_2}.\label{15}
\end{equation}
Numerically, the value of $T$ is determined mainly at the radius where $T_{\rm
merge}\simeq T_{\rm harden}$.

\section{Merger rate in GCs}\label{s4}

\subsection{Merger rate in a single GC}\label{s4.1}

To calculate the merger rate of IMRIs in an GC, besides the merger time $T$
calculated in the previous section, we also need to know the supply rate of
compact stellar objects, such as BHs and NSs, to the center where the IMBH
presumably resides. On one hand, if this rate is lower than the reciprocal of
the merger time ($1/T$), the formation of IMRIs would be limited by the supply
rate, and the merger rate, consequently, is equal to it as well.  On the other,
if the supply rate is higher than $1/T$, we expect two stellar compact objects
to arrive at the cluster center around the same time to form a triple system
with the IMBH. However, such a triple is almost always unstable and will likely
eject the lightest component. The result would be a tighter IMRI, which
continues to eject future stellar interlopers via dynamical hardening (see
previous section) until the IMRI coalesces. In this case, the average duration
between two successive mergers is limited by the merger time $T$, and,
consequently, the merger rate is $1/T$.

Following \citet{Enrico2}, we relate the supply rate of BHs and NSs to the
regrow timescale, $t_{\rm cusp}$, of the stellar cusp around an IMBH.  This
timescale depends on the size of the core scoured out by the hardening process
of the former IMRI and can be calculated approximately by
\begin{equation}
  t_{\rm cusp}\approx 6\left(\frac{M}{10^6M_\odot}\right)^{1.19}q^{0.35}\rm G yr,\label{14}
\end{equation}
where $M$ is the total mass of the IMRI and $q$ is the mass ratio between the
IMBH and the companion compact object \citep{Enrico}. Correspondingly, supply
rate is $1/(2t_{\rm cusp})$, where the factor of $2$ accounts for uncertainties
such as the time delay between cusp regeneration and IMRI formation.

\begin{figure}[!h]
\centering
\includegraphics[height=4.0in]{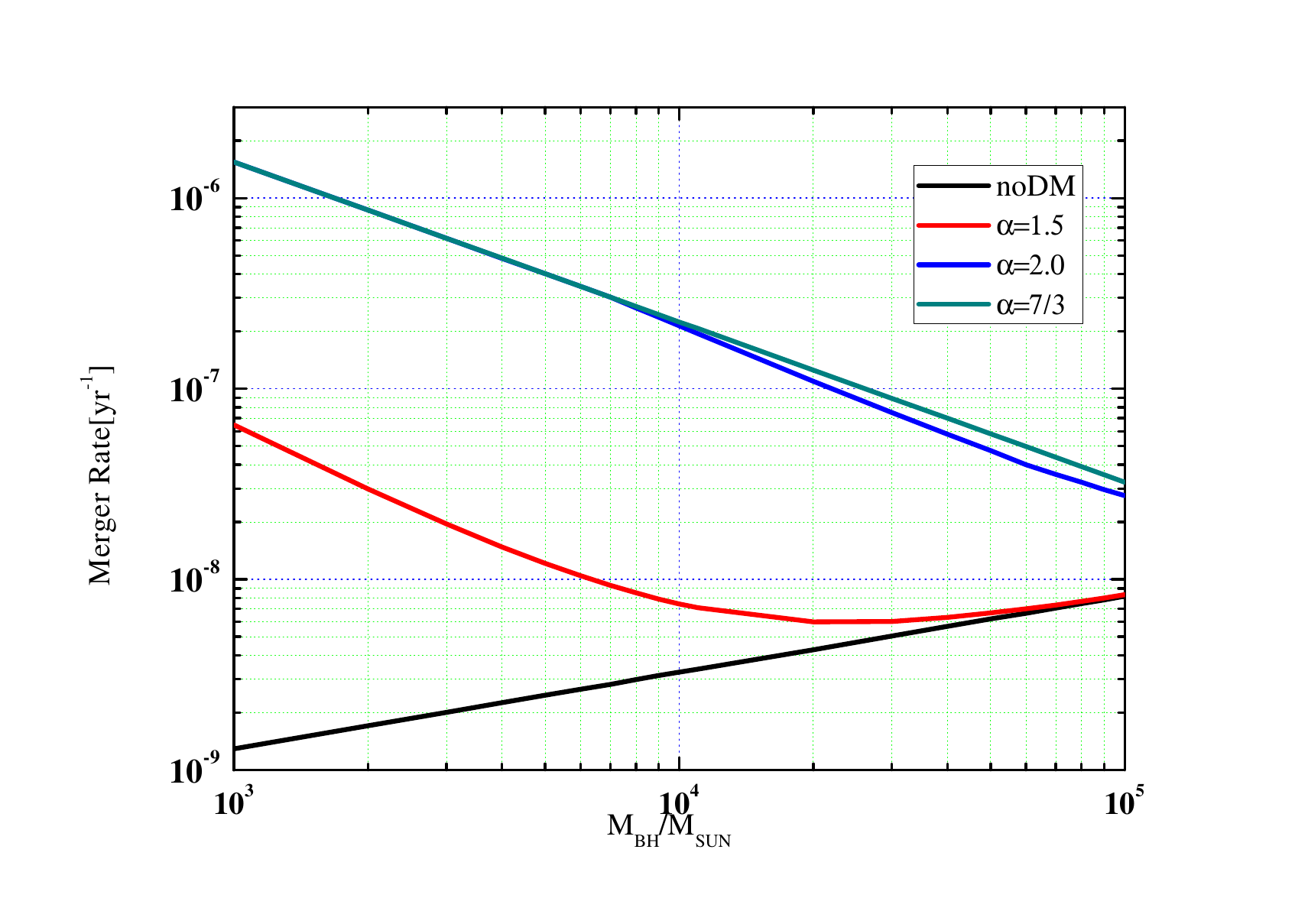}
\caption{Merger rate of BHs in a GC with different DM density profiles as a function
of the mass of the central IMBH.}
\label{fig04}
\end{figure}

Table~{\ref{table1}} summarizes our results for the calculation of the merger
rate.  It is clear that with DM minispikes, the merger rate is significantly
enhanced.  Figure~(\ref{fig04}) shows the merger rate of an IMBH with stellar BHs
as a function of the mass of the IMBH. We find that the presence of a DM
minispikes changes the relation between the merger rate and the IMBH mass.

\begin{table}[!h]
\centering
\caption{The reciprocal of the merger time, the supply rate of BHs to the vicinity of IMBHs to form IMRIs, the merger rate, and the duty cycle of IMRIs/EMRIs for different dark matter density profiles in a GC.} \label{table1}
\begin{tabular}{|c|c|c|c|c|}
\hline
$M=10^3 M_\odot$ & 1/T & Supply Rate & Merger Rate & Duty Cycle\\
\hline
No DM& $1.3\times 10^{-9}\textrm{yr}^{-1}$& $1.5\times 10^{-6}\textrm{yr}^{-1}$ & $1.3\times 10^{-9}\textrm{yr}^{-1}$ & $T_{\textrm{Hubble}}$\\
\hline
$\alpha=1.5$&$6.5\times 10^{-8}\textrm{yr}^{-1}$&$1.5\times 10^{-6}\textrm{yr}^{-1}$& $6.5\times 10^{-8}\textrm{yr}^{-1}$&$T_{\textrm{Hubble}}$ \\
\hline
$\alpha=2.0$&$2.1\times 10^{-6}\textrm{yr}^{-1}$&$1.5\times 10^{-6}\textrm{yr}^{-1}$& $1.5\times 10^{-6}\textrm{yr}^{-1}$& $6.7\times 10^8\textrm{yr}$  \\
\hline
$\alpha=7/3$&$6.6\times 10^{-6}\textrm{yr}^{-1}$&$1.5\times 10^{-6}\textrm{yr}^{-1}$& $1.5\times 10^{-6}\textrm{yr}^{-1}$& $6.7\times 10^8\textrm{yr}$ \\
\hline
\end{tabular}
\begin{tabular}{|c|c|c|c|c|}
\hline
$M=10^4 M_\odot$ & 1/T & Supply Rate & Merger Rate & Duty Cycle\\
\hline
No DM&$3.3\times10^{-9}\textrm{yr}^{-1}$&$2.2\times 10^{-7}\textrm{yr}^{-1}$& $3.3\times 10^{-9}\textrm{yr}^{-1}$ &$T_{\textrm{Hubble}}$\\
\hline
$\alpha=1.5$&$7.4\times 10^{-9}\textrm{yr}^{-1}$&$2.2\times 10^{-7}\textrm{yr}^{-1}$& $7.4\times 10^{-9}\textrm{yr}^{-1}$&$T_{\textrm{Hubble}}$ \\
\hline
$\alpha=2.0$&$2.1\times 10^{-7}\textrm{yr}^{-1}$&$2.2\times 10^{-7}\textrm{yr}^{-1}$& $2.1\times 10^{-7}\textrm{yr}^{-1}$& $4.8\times 10^9\textrm{yr}$\\
\hline
$\alpha=7/3$&$1.0\times 10^{-6}\textrm{yr}^{-1}$&$2.2\times 10^{-7}\textrm{yr}^{-1}$& $2.2\times 10^{-7}\textrm{yr}^{-1}$& $4.5\times 10^9\textrm{yr}$\\
\hline
\end{tabular}
\begin{tabular}{|c|c|c|c|c|c|}
\hline
$M=10^5M_\odot$ & 1/T & Supply Rate &  Merger Rate & Duty Cycle\\
\hline
No DM&$8.2\times 10^{-9}\textrm{yr}^{-1}$&$3.2\times 10^{-8}\textrm{yr}^{-1}$& $8.2\times 10^{-9}\textrm{yr}^{-1}$ & $T_{\textrm{Hubble}}$\\
\hline
$\alpha=1.5$&$8.3\times 10^{-9}\textrm{yr}^{-1}$&$3.2\times 10^{-8}\textrm{yr}^{-1}$& $8.3\times 10^{-9}\textrm{yr}^{-1}$ & $T_{\textrm{Hubble}}$ \\
\hline
$\alpha=2.0$&$2.7\times 10^{-8}\textrm{yr}^{-1}$&$3.2\times 10^{-8}\textrm{yr}^{-1}$&$2.7\times 10^{-8}\textrm{yr}^{-1}$& $T_{\textrm{Hubble}}$ \\
\hline
$\alpha=7/3$&$1.6\times 10^{-7}\textrm{yr}^{-1}$&$3.2\times 10^{-8}\textrm{yr}^{-1}$& $3.2\times 10^{-8}\textrm{yr}^{-1}$ & $T_{\textrm{Hubble}}$ \\
\hline
\end{tabular}
\end{table}

\begin{figure}[!h]
\end{figure}

Since the presence of DM minispikes can enhance the merger rate of IMBHs with
stellar BHs, a GC could be quickly deprived of stellar BHs and no longer
produce any IMRIs.  To quantify this probability, we calculate the duty cycles
for the production of IMRIs by our GCs, and the results are given in the last
column of Table~{\ref{table1}}. In our calculations we assume that each GC on
average has $1000$ stellar BHs. This number is derived based on the facts that
(1) a typical GC has $10^6$ stars, (2) about $0.3\%$ of them will turn into BHs
if the initial mass function (IMF) for stars is Salpeter and the progenitor
stars of BHs are more massive than $30M_\odot$,  and (3) a large fraction of
the newborn BHs will receive natal kicks that are greater than the escape
velocity of the GC. As a result, the GCs with a merger rate greater than about
$10^{-7}{\rm yr^{-1}}$ would deplete their stellar BHs within a Hubble time.
The relevant GCs, according to Figure~(\ref{fig04}), are those less massive
than $2\times10^4M_\odot$ and with a DM density profile of $\alpha\ge2.0$.

Besides depleting the stellar BHs in a cluster, another effect of the high
merger rate of IMRIs is to increase the mass of the IMBH on a timescale much
shorter than the Hubble time (see Table~{\ref{table1}). For example, when
$M_{\rm BH}=1000\,M_\odot$, the IMBH could grow to $2000M_\odot$ in less than
$10^8{\rm yrs}$ if $\alpha=7/3$, and after $5\,{\rm Gyrs}$ even to $\sim
10^4M_\odot$.  Such a rapid increment in mass would have a strong impact on the
structure of the star cluster around the IMBH and hence affect our calculations
of the merger rate and duty cycle, but our current model is not capable of
capturing this effect yet. Alternatively, if $\alpha<1.5$ or the mass is $M_{\rm
BH}>2\times10^4\,M_\odot$, the growth time would exceed the Hubble time and we can
neglect the accretion of BHs by IMBH in our model. The implication of these
results is that we could use IMRIs to constrain the mass function of IMBHs and
hence further test the DM models in GCs

After the exhaustion of BHs, the formation of IMRIs/EMRIs with NSs becomes
important. Table~(\ref{table2}) and Figure~(\ref{fig05}) show the merger rates
for NSs, which are calculated in a way similar to the BH merger rates but using
$\mu=1.4\,M_\odot$.  The maximum mass that we consider here for IMBH-NS mergers
is  $M_{\rm BH}=1\sim 3\times 10^4M_\odot$, because more massive IMBHs could
not deplete all the BHs in the host clusters.  The results for IMBH-NS
mergers should be taken with caution because we calculated the hardening rate
using Equation~(\ref{12}), which is derived under the assumption that the
intruding stars are less massive than the smaller member of the IMRI/EMRI. This
assumption may be invalid when the IMRI/EMRI involves an NS. With a
Salpeter mass function and the assumption that stars within $10M_\odot$ and
$30M_\odot$ will form NSs, we find that NSs in GCs will not be exhausted in our
models.

\begin{table}[!h]
\caption{The merger rate of IMRIs or EMRIs with neutron stars for different dark matter density profiles in a GC.} \label{table2}
\begin{center}
\begin{tabular}{|c|c|c|c|c|}
\hline
  & $M=10^4M_\odot$ & $M=2\times 10^4M_\odot$ & $M=3\times 10^4M_\odot$\\
\hline
$\alpha=2.0$& $5.9\times 10^{-8}\rm yr^{-1}$& $3.1\times 10^{-8}\rm yr^{-1}$ & $2.2\times10^{-8}\rm yr^{-1}$  \\
\hline
$\alpha=7/3$& $3.6\times 10^{-7}\rm yr^{-1}$& $2.0\times10^{-7}\rm yr^{-1}$ & $1.5\times10^{-7}\rm yr^{-1}$  \\
\hline
\end{tabular}
\end{center}
\end{table}

\begin{figure}[!h]
\centering
\includegraphics[height=4.0in]{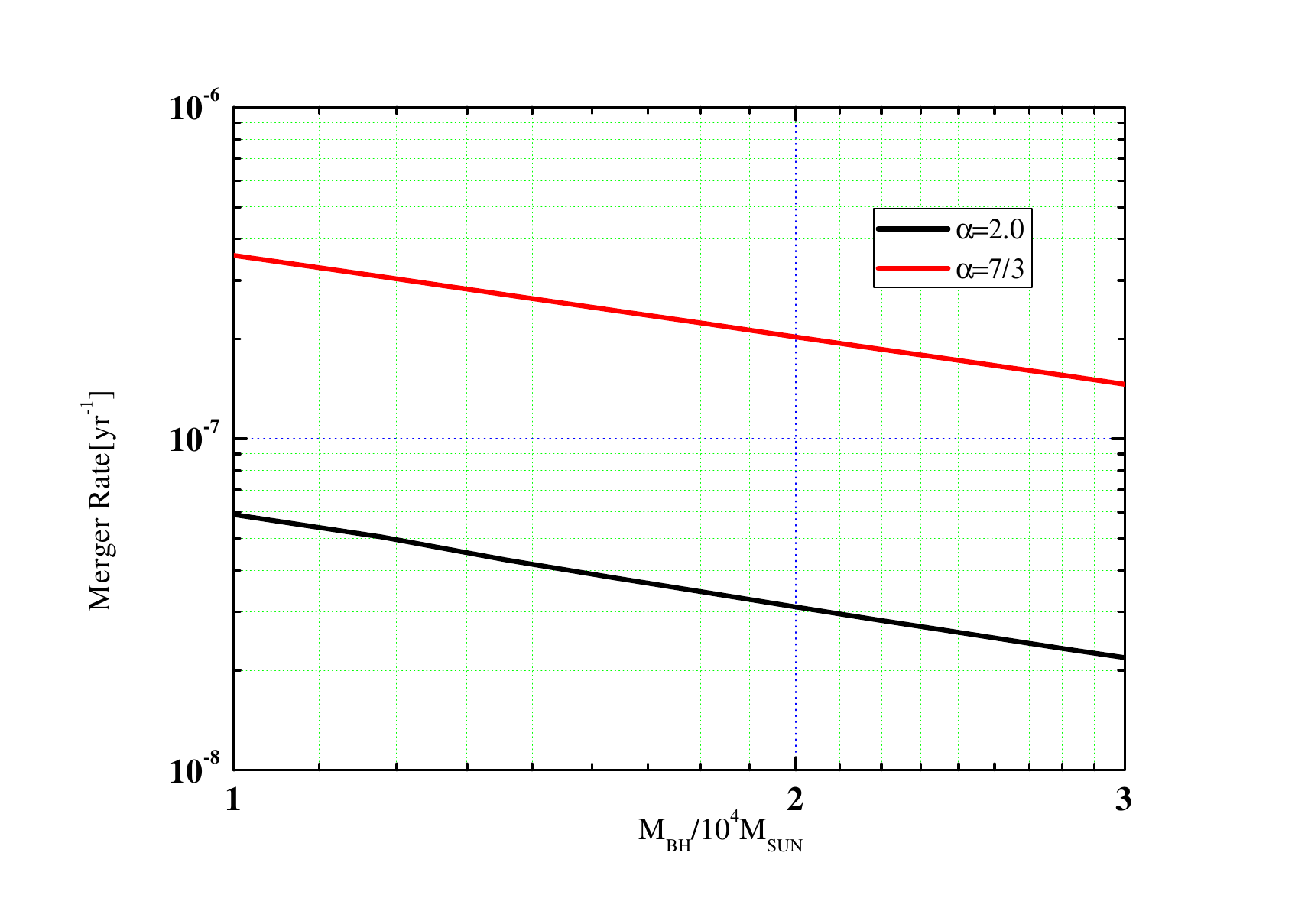}
\caption{Same as Figure~\ref{fig04} but for IMBH-NS mergers in GCs.}
\label{fig05}
\end{figure}

Having derived the merger rate for BHs and NSs, we can study the growth of IMBH
during one Hubble time. The results are shown in Figure~\ref{fig06} as a
function of the initial masses of the IMBHs. We find that in the presence of DM
minispikes, the growth of IMBHs cannot be neglected when the intial masses are
small.  Alternatively, when there is no minispike, the growth of IMBHs due to
the mergers with BHs  can be neglected.

\begin{figure}[!h]
\centering
\includegraphics[height=4.0in]{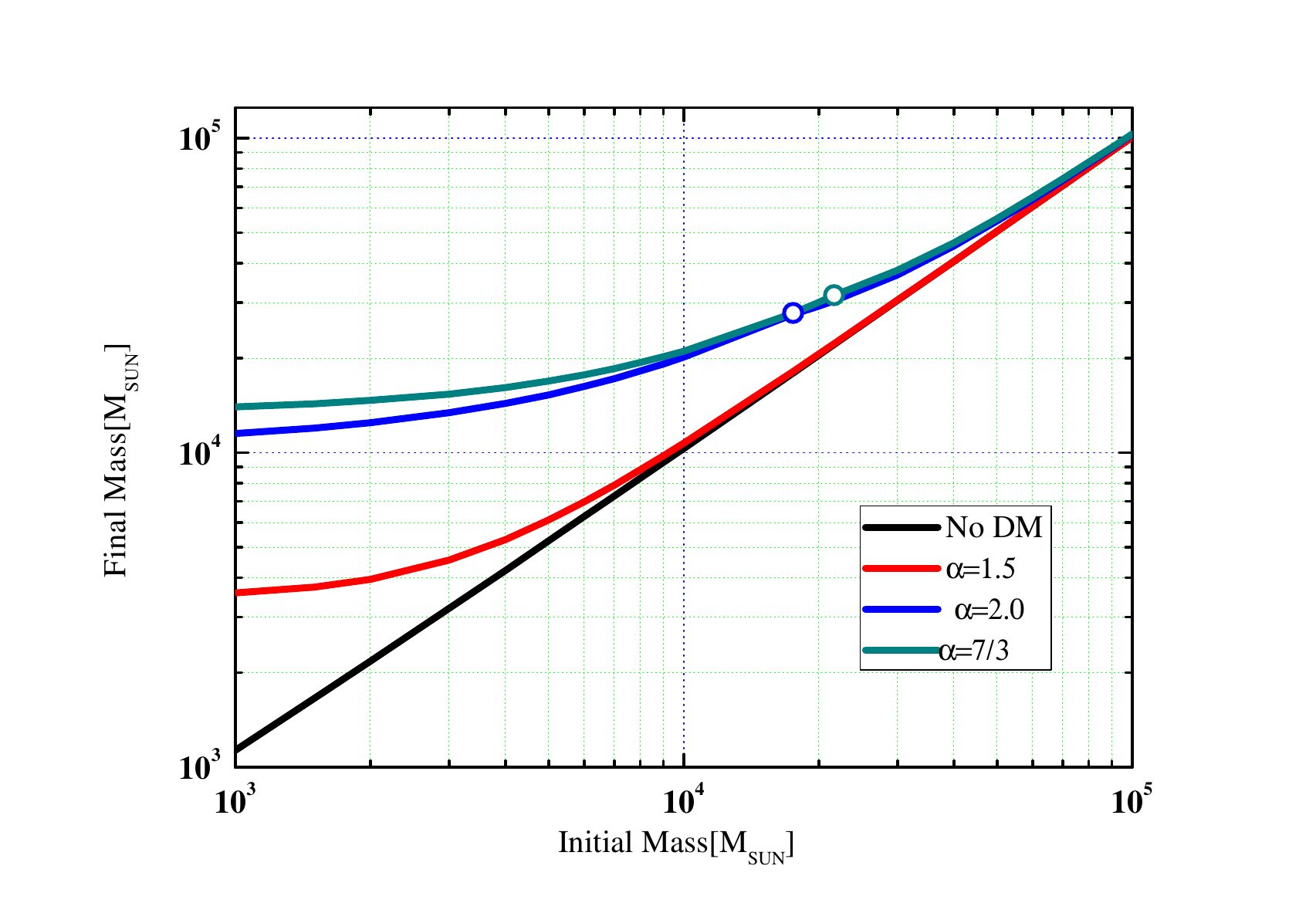}
\caption{Final masses of IMBHs after 10 Gyr for different initial masses of IMBHs and density profiles of DM cusps in GCs. The circles mark the critical masses below which the
IMBHs cannot deplete all the stellar BHs in the clusters. The corresponding
initial masses are about $17,600M_\odot$ for
$\alpha=2.0$ and $21,700M_\odot$ for $\alpha=7/3$.}
\label{fig06}
\end{figure}

\subsection{Merger rate in GCs per $\rm Mpc^3$}\label{s4.2}

To derive the merger rate of IMRIs/EMRIs per unit volume, we first calculate
the number of GCs in a single galaxy. We adopt the conventional assumption that
a fraction of $f_{\rm gc}=0.05$ of stars are in clusters \citep{Kruijssen,Gnedin},
and the mass function of GCs follows a power law $dn/dm_{\rm gc}\propto
m_{\rm gc}^{-2}$, where $m_{\rm gc}$ is the cluster mass whose limits are $m_{\rm
min}=10^2M_\odot$ and $m_{\rm max}=10^7M_\odot$ \citep{Bik,de Grijs}. Then
given the mass of a galaxy, $\mathcal{M}$, the GC mass function is
\begin{equation}
 \frac{dn}{dm_{\rm gc}}=\frac{0.05\mathcal{M}}{\ln( 10^5)m_{\rm gc}^2}.\label{15}
\end{equation}

Second, we calculate the number of galaxies per unit volume.
According to \citet{Baldry}, the field galaxy stellar-mass function is
\begin{equation}
\phi_{\mathcal{M}}d\mathcal{M}=e^{-\mathcal{M}/\mathcal{M}_\ast}
\left[\phi_1^\ast\left(\frac{\mathcal{M}}{\mathcal{M}_\ast}\right)^{\alpha_1} +\phi_2^\ast\left(\frac{\mathcal{M}}{\mathcal{M}_\ast}\right)^{\alpha_2} \right]\frac{d\mathcal{M}}{\mathcal{M}_\ast},\label{15}
\end{equation}
where $\phi_{\mathcal{M}}d\mathcal{M}$ is the number density of galaxies within
the mass bin between $\mathcal{M}$ and $\mathcal{M}+d\mathcal{M}$.  The best-
fitting parameters are
\begin{eqnarray}
\log(\mathcal{M}/M_\odot)&&=10.648\nonumber\\
\phi_1^\ast/(10^{-3}\rm Mpc^{-3})&&=4.26,\alpha_1=-0.46\nonumber\\
\phi_2^\ast/(10^{-3}\rm Mpc^{-3})&&=0.58,\alpha_2=-1.58.\label{16}
\end{eqnarray}

Third, we derive the number density of GCs as
\begin{equation}
n= \int_{\mathcal{M}_{\rm min}}^{\mathcal{M}_{\rm max}}\left(\int_{m_{\rm min}}^{m_{\rm max}}\frac{dn}{dm_{\rm gc}}dm_{\rm gc}\right)\phi_{\mathcal{M}}d\mathcal{M},\label{17}
 \end{equation}
where we choose the integral limits for galaxies to be
$\mathcal{M}_{\rm min}=10^8M_\odot$ and $\mathcal{M}_{\rm max}=10^{12}M_\odot$.
 If we assume
IMBHs exist in the GCs whose masses exceed $10^5\,M_\odot$, we find that a space
density of $n=9.4\,{\rm Mpc^{-3}}$ for IMBHs.

As for the mass function of IMBHs, it is poorly observationally constrained.
Here we assume a power-law function,
\begin{equation}
\frac{dn}{dM}\propto M^\beta,\label{18}
\end{equation}
with the index $\beta$ a free parameter.  In the later calculations, we consider
different values for $\beta$, namely, $-2.0$, $-1.0$, and $-0.5$.  The merger
rates for BHs and NSs, which we refer to as $\mathcal{R}_{\rm BH}(M)$ and
$\mathcal{R}_{\rm NS}(M)$, have been derived in Figures~\ref{fig04} and
\ref{fig05}.

Finally, we can calculate the merger rate density using the quantities
defined above. When there is no DM minispike, we simply have
\begin{equation}
  R_{\rm BH}=\int_{1000M_\odot}^{10^5M_\odot}\mathcal{R}_{\rm BH}(M)\frac{dn}{dM}dM.\label{19}
\end{equation}
Alternatively, if DM minispikes exist,  the difference is that the enhanced
merger rate could modify the IMBH mass function as we have seen in
Section~\ref{s4.1}. In this case, we convert the initial mass of an IMBH
into the final mass $M_f=f(M_i)$ according to
Figure~(\ref{fig06}), and calculate the merger rate of BHs with
\begin{equation}
     R_{\rm BH}=\int_{M_{\rm min}}^{M_{\rm max}}\mathcal{R}_{\rm BH}(f(M))\frac{dn}{dM}dM.\label{20}
\end{equation}
In the above integration, the maximum mass is chosen to be
$M_{\rm max}$=$10^5M_\odot$. The minimum mass is $M_{\rm min}=1000M_\odot$ when
$\alpha=1.5$, but when $\alpha=2.0$ and $7/3$, it is chosen to be the maximum
mass within which an IMBH could deplete all stellar BHs in the host cluster
during one Hubble time. For IMBHs that are below this mass limit,
the mergers with NSs become predominant. In this case, we calculate the NS merger rate
as
\begin{equation}
     R_{\rm NS}=\int_{M_{\rm min}}^{M_{\rm max}}\mathcal{R}_{\rm NS}(f(M))\frac{dn}{dM}dM,\label{201}
\end{equation}
where $M_{\rm min}$ is $1000M_\odot$ and $M_{\rm max}$ should be the mass that
allows IMBHs to deplete all BHs.

\begin{table}[!h]
\caption{The merger rate in GCs per $\rm Mpc^3$ of IMRIs or EMRIs with BHs $R_{\rm BH}$ and neutron stars $R_{\rm NS}$  for different dark matter density profiles. } \label{table3}
\begin{center}
\begin{tabular}{|c|c|c|c|c|}
\hline
$\beta=-2.0$ & No DM & $\alpha=1.5$ & $\alpha=2.0$ & $\alpha=7/3$\\
\hline
$R_{\rm BH}[\rm Mpc^{-3} yr^{-1}$]& $1.9\times10^{-8}$& $1.3\times 10^{-7}$ & $2.7\times 10^{-8}$ & $2.2\times 10^{-8}$ \\
\hline
$R_{\rm NS}[\rm Mpc^{-3} yr^{-1}$]& -- & -- & $4.1\times10^{-7}$ & $2.3\times10^{-6}$ \\
\hline
\end{tabular}
\end{center}

\begin{center}
\begin{tabular}{|c|c|c|c|c|}
\hline
$\beta=-1.0$ & No DM & $\alpha=1.5$ & $\alpha=2.0$ & $\alpha=7/3$\\
\hline
$R_{\rm BH}[\rm Mpc^{-3} y^{-1}$]& $3.5\times10^{-8}$& $8.9\times 10^{-8}$ & $1.8\times 10^{-7}$ & $1.8\times 10^{-7}$ \\
\hline
$R_{\rm NS}[\rm Mpc^{-3} yr^{-1}$]& -- & -- & $2.4\times10^{-7}$ & $1.4\times10^{-6}$ \\
\hline
\end{tabular}
\end{center}

\begin{center}
\begin{tabular}{|c|c|c|c|c|}
\hline
$\beta=-0.5$ & No DM & $\alpha=1.5$ & $\alpha=2.0$ & $\alpha=7/3$\\
\hline
$R_{\rm BH}[\rm Mpc^{-3} yr^{-1}$]& $4.6\times10^{-8}$& $7.4\times 10^{-8}$ & $2.9\times 10^{-7}$ & $3.0\times 10^{-7}$ \\
\hline
$R_{\rm NS}[\rm Mpc^{-3} yr^{-1}$]& -- & -- & $1.2\times10^{-7}$ & $7.9\times10^{-7}$ \\
\hline
\end{tabular}
\end{center}

\end{table}

Table~\ref{table3} shows the resulting merger rates for BHs and NSs.  We find
that when $\alpha\ge2.0$, the merger rate changes more significantly as the
$\beta$ parameter varies. This is because the DM minispikes start to play a
role in determining the merger rate by depleting the stellar BHs in the same
cluster. {It is easy to estimate event rates for LISA based on these merger rate results. 
Conservatively, if LISA can detect IMRIs as far as 1 Gpc, LISA will see $\sim 10^2$ IMRIs per Gpc$^3$ every year. This high rate is due to the assumptions of the presence of a DM minispike and every GC having an IMBH.}

{When $\beta=-2.0$, we can find an interesting phenomena in which the BH
merger rate decreases for steeper DM cusps. This is due to the exhaustion of BH
supply.  When $\alpha=1.5$, BHs are not exhausted, but as $\alpha$ increases to
$2.0$ and $7/3$, BHs are exhausted in the GCs with low-mass IMBHs.  As a result,
the population of IMRIs becomes smaller when $\alpha$ increases, and the merger
rate per unit ${\rm Mpc}^3$, correspondingly, decreases.}

\section{Merger rate in nuclear
star clusters}\label{s5} \subsection{Merger rate in one NSC}\label{s5.1}

Our method of calculating the IMRI/EMRI rates in NCSs is essentially the same as
the one presented in the previous section, except that the physical parameters
for NSCs are different. In particular, the velocity dispersion of an NSC and the
mass of the central IMBH satisfies the $M_{\rm BH}-\sigma$ relation
\begin{equation}
\textrm{log}\left(\frac{M_{\rm BH}}{M_{\odot}}\right)=8.12+4.24\textrm{log}\left(\frac{\sigma}{200\textrm{km/s}}\right)\label{21}
\end{equation}
\citep[also see][]{Gurkan}. The density is much higher than that in GCs and
observations suggest that it is  $10^{6-7}M_\odot\,{\rm pc^{-3}}$
\citep{Phillips,Walcher}.  In the following calculations, we use the lower
limit $10^6M_\odot\,{\rm pc^{-3}}$ to derive conservative values for the merger
rates.

Table~\ref{table4} summarizes the results of our calculations of the reciprocal
of the merger time due to dynamical friction and GW radiation, the supply rate
of BHs due to stellar relaxation, and the final merger rate.  Moreover, the
final merger rate is illustrated in Figure~\ref{fig07}. Now the merger rates
for $\alpha=2.0$ and $\alpha=7/3$ are the same because both are determined by
the supply rate.

\begin{table}[!h]
\centering
\caption{The same as Table~\ref{table1} but for NSCs.} \label{table4}
\begin{tabular}{|c|c|c|c|c|}
\hline
$M=10^3 M_\odot$ & 1/T & Supply Rate & Merger Rate & Duty Cycle\\
\hline
No DM& $4.9\times 10^{-9}\textrm{yr}^{-1}$& $1.5\times 10^{-6}\textrm{yr}^{-1}$ & $4.9\times 10^{-9}\textrm{yr}^{-1}$ & $T_{\textrm{Hubble}}$\\
\hline
$\alpha=1.5$&$8.5\times 10^{-8}\textrm{yr}^{-1}$&$1.5\times 10^{-6}\textrm{yr}^{-1}$& $8.5\times 10^{-8}\textrm{yr}^{-1}$&$2.4\textrm{Gyr}$ \\
\hline
$\alpha=2.0$&$3.7\times 10^{-6}\textrm{yr}^{-1}$&$1.5\times 10^{-6}\textrm{yr}^{-1}$& $1.5\times 10^{-6}\textrm{yr}^{-1}$& $1.3\times 10^8\textrm{yr}$  \\
\hline
$\alpha=7/3$&$1.4\times 10^{-5}\textrm{yr}^{-1}$&$1.5\times 10^{-6}\textrm{yr}^{-1}$& $1.5\times 10^{-6}\textrm{yr}^{-1}$& $1.3\times 10^8\textrm{yr}$ \\
\hline
\end{tabular}
\begin{tabular}{|c|c|c|c|c|}
\hline
$M=10^4 M_\odot$ & 1/T & Supply Rate & Merger Rate & Duty Cycle\\
\hline
No DM&$8.0\times10^{-9}\textrm{yr}^{-1}$&$2.2\times 10^{-7}\textrm{yr}^{-1}$& $3.3\times 10^{-9}\textrm{yr}^{-1}$ &$T_{\textrm{Hubble}}$\\
\hline
$\alpha=1.5$&$1.3\times 10^{-8}\textrm{yr}^{-1}$&$2.2\times 10^{-7}\textrm{yr}^{-1}$& $1.3\times 10^{-8}\textrm{yr}^{-1}$&$T_{\textrm{Hubble}}$ \\
\hline
$\alpha=2.0$&$3.2\times 10^{-7}\textrm{yr}^{-1}$&$2.2\times 10^{-7}\textrm{yr}^{-1}$& $2.2\times 10^{-7}\textrm{yr}^{-1}$& $9.1\textrm{Gy}$\\
\hline
$\alpha=7/3$&$1.7\times 10^{-6}\textrm{yr}^{-1}$&$2.2\times 10^{-7}\textrm{yr}^{-1}$& $2.2\times 10^{-7}\textrm{yr}^{-1}$& $9.1\textrm{Gy}$\\
\hline
\end{tabular}
\begin{tabular}{|c|c|c|c|c|c|}
\hline
$M=10^5M_\odot$ & 1/T & Supply Rate &  Merger Rate & Duty Cycle\\
\hline
No DM&$1.3\times 10^{-8}\textrm{yr}^{-1}$&$3.2\times 10^{-8}\textrm{yr}^{-1}$& $1.3\times 10^{-8}\textrm{yr}^{-1}$ & $T_{\textrm{Hubble}}$\\
\hline
$\alpha=1.5$&$1.3\times 10^{-8}\textrm{yr}^{-1}$&$3.2\times 10^{-8}\textrm{yr}^{-1}$& $1.3\times 10^{-8}\textrm{yr}^{-1}$ & $T_{\textrm{Hubble}}$ \\
\hline
$\alpha=2.0$&$3.6\times 10^{-8}\textrm{yr}^{-1}$&$3.2\times 10^{-8}\textrm{yr}^{-1}$&$3.2\times 10^{-8}\textrm{yr}^{-1}$& $T_{\textrm{Hubble}}$ \\
\hline
$\alpha=7/3$&$2.1\times 10^{-7}\textrm{yr}^{-1}$&$3.2\times 10^{-8}\textrm{yr}^{-1}$& $3.2\times 10^{-8}\textrm{yr}^{-1}$ & $T_{\textrm{Hubble}}$ \\
\hline
\end{tabular}
\end{table}

\begin{figure}[!h]
\centering
\includegraphics[height=4.0in]{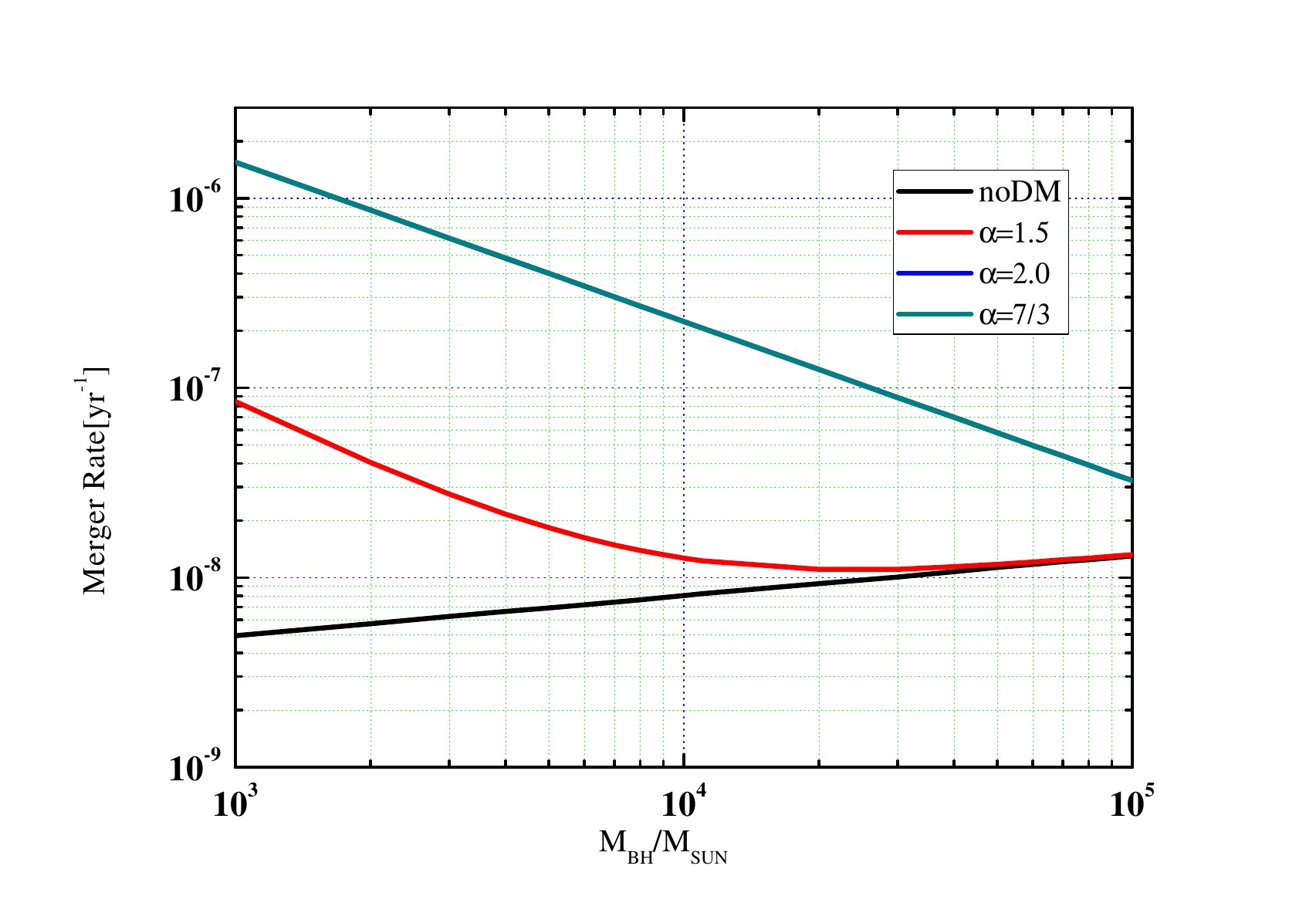}
\caption{Same as Figure~\ref{fig05} but for NSCs.}
\label{fig07}
\end{figure}

As for the merger rate of GCs, we first need to assess the number of BHs in
NSCs and to estimate the corresponding depletion timescale. We note that the total
mass of an NSC is typically $100$ times that of the central IMBH \citep[see Fig.
5 of][]{Antonini}. Following our earlier assumption that the average stellar-
mass is $0.5M_\odot$, the stellar IMF is a Salpeter function, and only one-third of
the newborn stellar BHs are retained, we find that the total number of BHs is
$200$ for $M=1000M_\odot$, $2000$ for $M=10,000M_\odot$, and $20,000$ for
$M=10^5M_\odot$.

With these numbers and not considering the effect of mass growth, we derive the
duty cycles of BH mergers in NSCs, and the results are presented in
Table~\ref{table4}. We find that the duty cycles are now comparable to one
Hubble time when $M_{\rm BH}\ge10^4M_\odot$. If we further consider the growth
of IMBH due to the mergers, the duty cycles would be even longer because they
are  decreasing functions of $M_{\rm BH}$.

For those models that have a duty cycle shorter than one Hubble time, we need
to consider BH depletion and the subsequent NS mergers.  The merger rates are
given in Table.(\ref{table5}) and illustrated in Figure~\ref{fig08}. Because we
find that NS merger rate could be much higher than that in GCs, we also need to
consider the possibility of NS depletion.  We estimate the total number of NSs
in an NSC assuming that the stars in the mass range between $10M_\odot$ and
$30M_\odot$ will turn into NSs.  As a result, about $1$ percent of the initial
stars would form NSs. Moreover, we assume that half the NSs, due to their large
natal kicks,  would escape from the NSCs.  Then we find about $1000$ NSs for an
IMBH of an initial mass $M_{\rm BH}=1000M_\odot$, $10^4$ for $M_{\rm BH}=10^4M_\odot$ and $10^5$
for $M_{\rm BH}=10^5M_\odot$. The resulting NS duty cycles are presented in
Table~\ref{table5}.

\begin{table}[!h]
\caption{The merger rate of IMRIs/EMRIs with NSs for different dark matter density profiles in NSCs.} \label{table5}
\begin{center}
\begin{tabular}{|c|c|c|c|}
\hline
  $M=3000M_\odot$ &$\alpha=1.5$ & $\alpha=2.0$ & $\alpha=7/3$\\
\hline
Merger Rate& $8.0\times 10^{-9}\rm yr^{-1}$& $3.1\times 10^{-7}\rm yr^{-1}$ & $1.2\times 10^{-6}$  \\
\hline
Duty Cycle& $T_{\rm Hubble}$& $3.2\rm Gyr$ & $8.3\times10^8\rm yr$  \\
\hline
\end{tabular}

\begin{tabular}{|c|c|c|c|}
\hline
  $M=12,000M_\odot$ &$\alpha=1.5$ & $\alpha=2.0$ & $\alpha=7/3$\\
\hline
Merger Rate& $6.2\times 10^{-9}\rm yr^{-1}$& $7.3\times 10^{-8}\rm yr^{-1}$ & $3.8\times 10^{-7}\rm yr^{-1}$  \\
\hline
Duty Cycle& $T_{\rm Hubble}$ & $T_{\textrm{Hubble}}$ & $T_{\rm Hubble}$  \\
\hline
\end{tabular}
\begin{tabular}{|c|c|c|c|}
\hline
  $M=21,000M_\odot$ &$\alpha=1.5$ & $\alpha=2.0$ & $\alpha=7/3$\\
\hline
Merger rate& $6.6\times 10^{-9}\rm yr^{-1}$& $4.3\times 10^{-8}\rm yr^{-1}$ & $2.4\times 10^{-7}\rm yr^{-1}$  \\
\hline
duty cycle& $T_{\rm Hubble}$& $T_{\rm Hubble}$ & $T_{\rm Hubble}$  \\
\hline

\end{tabular}

\end{center}
\end{table}

\begin{figure}[!h]
\centering
\includegraphics[height=4.0in]{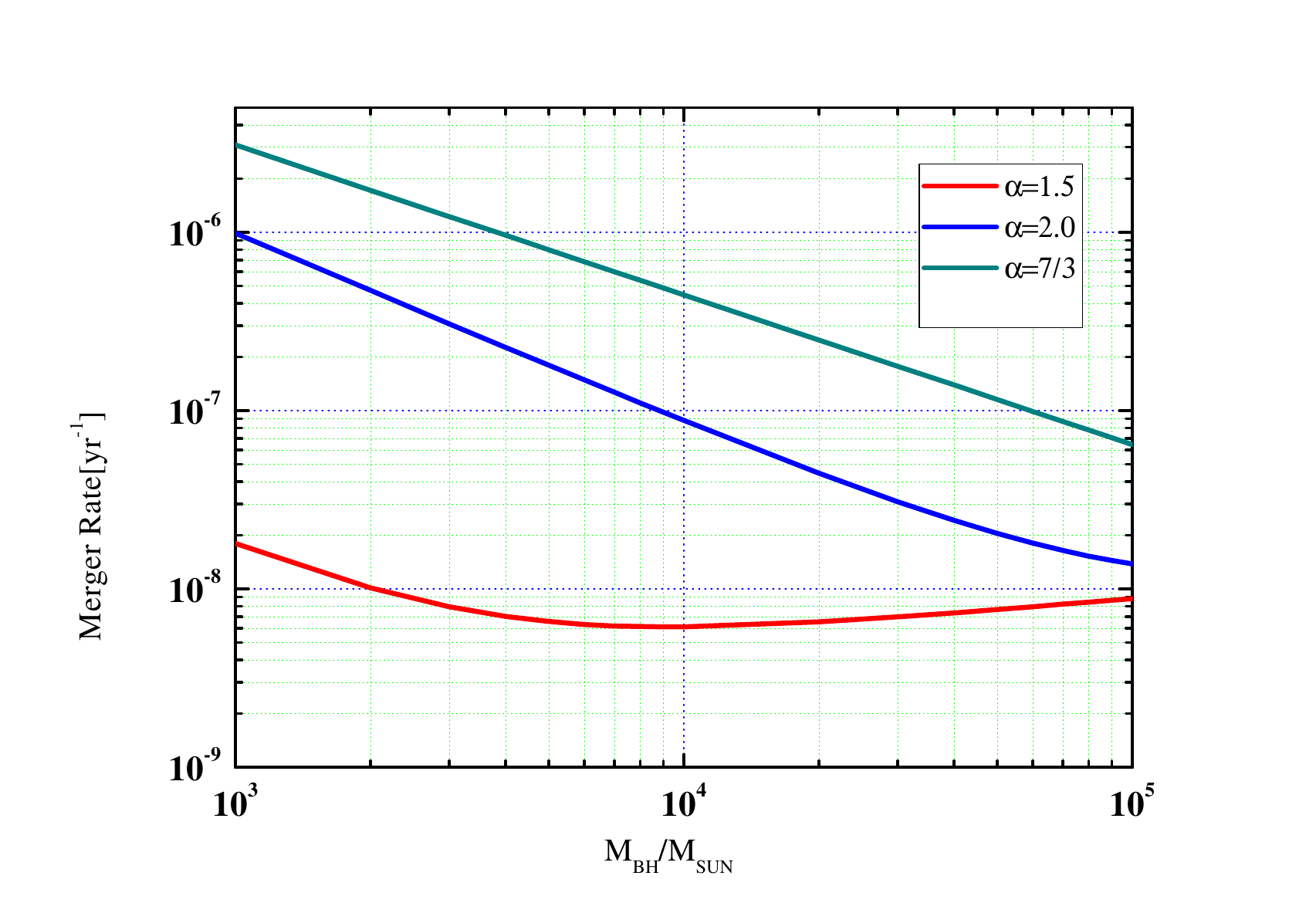}
\caption{Same as Figure~\ref{fig05} but for IMBH-NS merger NSCs.}
\label{fig08}
\end{figure}

Figure~\ref{fig09} shows the final masses of IMBHs as a function of the initial
ones.  For those IMBHs that could deplete NSs during one Hubble time, the
final masses include the contribution from all the stellar BHs and NSs in their
host NSCs.  \comment{without the addition of the  following  accretion
process.}

\begin{figure}[!h]
\centering
\includegraphics[height=4.0in]{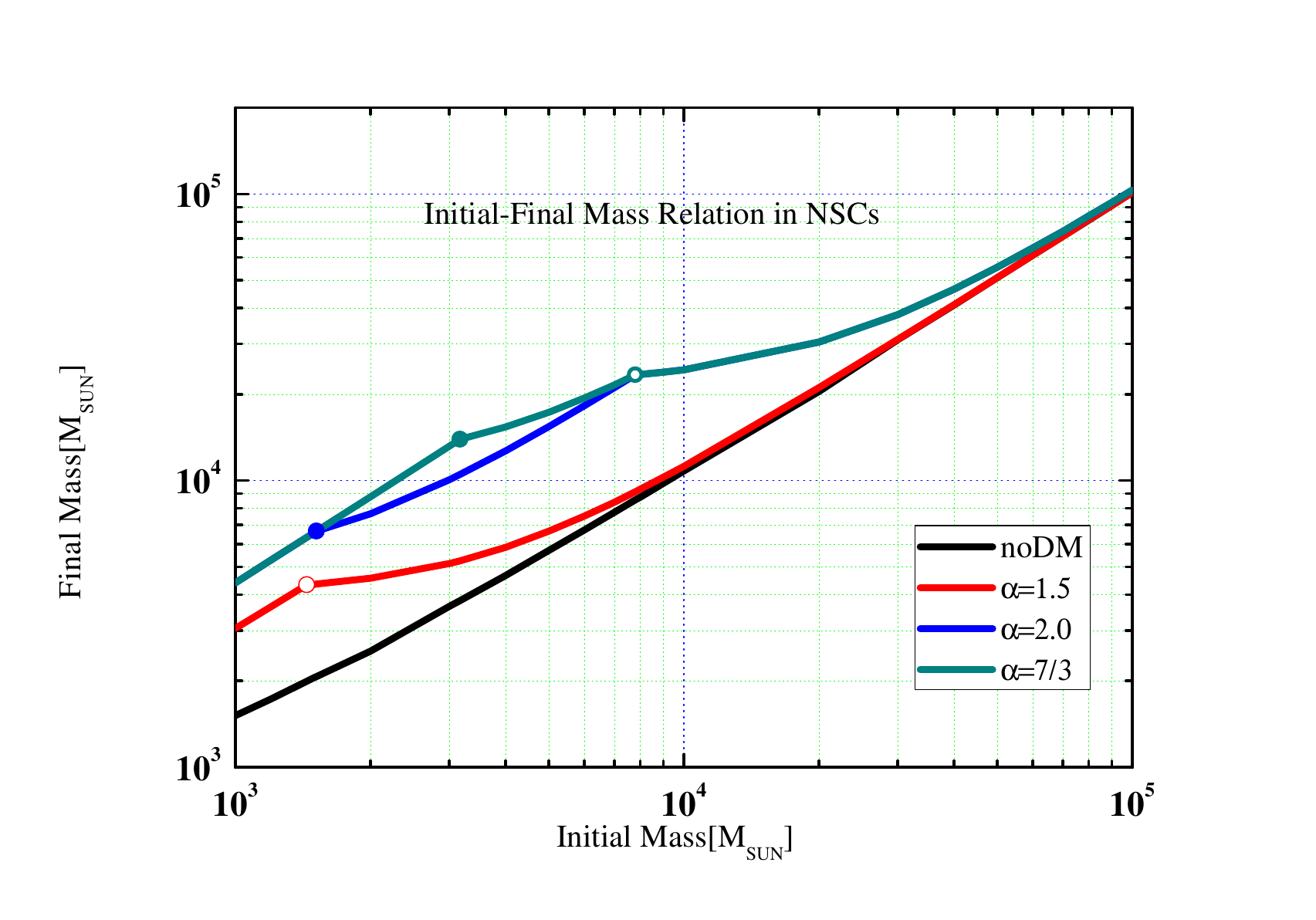}
\caption{Same as Figure~\ref{fig06} but for the IMBHs in NSCs. The circles,
at  about $1500M_\odot$ when $\alpha=1.5$ and $7800$ when $\alpha=2.0$ and
$7/3$, mark the masses with which a IMBH could deplete all the stellar BH in
the same cluster.  The solid dots are the masses corresponding to NS depletion,
which is about $ 1500M_\odot$ when $\alpha=2.0$ and $3200M_\odot$ when
$\alpha=7/3$. }
\label{fig09}
\end{figure}

\comment{\textcolor{red}{When BHs and NSs are consumed,   the final mass of
IMBHs is about $4000M_\odot<M<15000M_\odot$. the following accretions are white
dwarfs and the remaining main sequence stars. Whether they can be disrupted
before merger depends on the comparison of the tidal radius $r_{\rm tide}$ and
the innermost stable radius $r_{\rm ISCO}$.  The $r_{\rm tide}$ was calculated
to be $2.5(M/\mu)^{1/3}R$, where $R$ and $\mu$ is the radius and mass of the
inspiralling object \citep{I11}. For white dwarfs, $\mu \approx 0.6M_\odot$ and
$R\approx 10^7{\textrm {m}}$.  If we choose $M=10^4M_\odot$, $r_{\rm tide}\sim
6\times 10^8\rm m$, which is larger than $r_{\rm ISCO}$ and a tidal disruption
event(TDE) occurs. However, at such a separation the time to merge only via GW
is  about $0.1 \rm yr$, which makes it hard to observe.}}

\comment{\textcolor{red}{With a remaining main sequence star of $0.6M_\odot$,
the radius is $40$ times larger than the white dwarf. The separation of
disruption is $40$ times higher and the merger time is $40^4$ times greater,
which is $\sim 3\times 10^5\rm yr$, and the DM minispike may still make this
time much shorter.  Whether TDEs like this can be observed requires more
discussions in the future.}}

\subsection{Merger rate in nuclear star clusters per $\rm Mpc^3$}\label{s5.2}

The calculation of the merger rate per unit ${\rm Mpc^3}$ for NSCs is almost
the same as the previous calculation for GCs, but with two crucial
modifications.  (1) For the mass function of IMBHs, we consider two models.
The first is suggested by \citet{Enrico}, which is
 \begin{equation} \frac{{\rm d} n}{{\rm d}\log
M}=0.005\left(\frac{M}{3\times10^6M_\odot}\right)^{-0.3}\rm Mpc^{-3}.\label{22}
\end{equation}
Another model, suggested by \citet{Enrico2}, is more conservative,
\begin{equation} \frac{{\rm d} n}{{\rm d}\log
M}=0.002\left(\frac{M}{3\times10^6M_\odot}\right)^{0.3}\rm Mpc^{-3}.\label{23}
\end{equation}
In this latter one, the exponent is positive, which significantly reduces the
number of IMBHs. (2) When $\alpha\ge2.0$, to calculate $\mathcal{R}_{\rm NS}$ the $M_{\rm min}$ in the integration
of the mass function is determined by the solid dots in Figure~\ref{fig09},
because smaller IMBHs in NSCs would have depleted all the NSs.  The
corresponding merger rates for BHs and NSs per $\rm Mpc^3$ are presented in
Table\ref{table6}. Again, we find an enhancement of BH and NS merger rates
in the presence of DM minispikes.

 \begin{table}[!h]
\caption{The merger rate per $\rm Mpc^3$ of IMRIs or EMRIs with BHs $R_{\rm BH}$ and neutron stars $R_{\rm NS}$  formed by IMBH for different dark matter density profiles. } \label{table6}
\begin{center}
\begin{tabular}{|c|c|c|c|c|}
\hline
Barrause 12 & no DM & $\alpha=1.5$ & $\alpha=2.0$ & $\alpha=7/3$\\
\hline
$R_{\rm BH}[\rm Mpc^{-3} {\rm yr}^{-1}$]& $1.0\times10^{-9}$& $1.8\times 10^{-9}$ & $4.3\times 10^{-9}$ & $4.3\times 10^{-9}$ \\
\hline
$R_{\rm NS}[\rm Mpc^{-3} {\rm yr}^{-1}$]& -- & $1.4\times10^{-10}$ & $5.4\times10^{-9}$ & $8.7\times10^{-9}$ \\
\hline
\end{tabular}
\end{center}

\begin{center}
\begin{tabular}{|c|c|c|c|c|}
\hline
Gair 10 & no DM & $\alpha=1.5$ & $\alpha=2.0$ & $\alpha=7/3$\\
\hline
$R_{\rm BH}[\rm Mpc^{-3} {\rm yr}^{-1}$]& $1.7\times10^{-11}$& $2.3\times 10^{-11}$ & $8.9\times 10^{-11}$ & $8.9\times 10^{-11}$ \\
\hline
$R_{\rm NS}[\rm Mpc^{-3} {\rm yr}^{-1}$]& -- & $5.1\times10^{-13}$ & $3.4\times 10^{-11}$ & $7.4\times10^{-11}$ \\
\hline
\end{tabular}
\end{center}
\end{table}

From the calculations we can also find that  the results are sensitive to the BH and NS supply. With DM minispike, the duty cycle of BHs and NSs may be shorter than the Hubble time, as the dramatically increased merger rate can  lead to a high efficiency of consumption of compact objects. As a result, the appearance nowadays strongly depends on the total BH and NS supply.

\section{Could stellar dynamics deplete the minispikes?}\label{s6}

We have shown in Figures~\ref{fig01}-\ref{fig03} that dynamical
hardening, by ejecting intruding stars, is important for the initial evolution
of IMRI/EMRIs, when $a>(10^{-3}\sim0.1)$ pc.  It is worth discussing here
whether the same process could eject DM particles and hence deplete the DM
minispikes.

 The difference between hardening against DM and hardening
against stars is that the interloper stars that we considered are
gravitationally unbound to the IMBHs but the DM particles are deep in the
potential well of the IMBH. The DM particles are more difficult to deplete. To
show this more clearly, we calculate the hardening timescale associated with DM
using the ejection timescale derived in \citet{Sesana} for the stars
gravitationally bound to binary MBHs. They showed that it is $\tau=5P_0/q^2$,
where $P_0$ is the orbital period of the binary. For example, the pink solid
line in Figure~\ref{fig10} illustrates the dependence of this timescale on the
IMBH mass.

The fact that the pink line is comparable or above the blue and
green curves suggests that dynamical friction against DM is more efficient, so
that the shrinking of the binary does not lead to the ejection of most DM
particles, at least when $\alpha\ge2$. In the case where $\alpha=1.5$, the
hardening due to DM could be more efficient than the dynamical friction
process, which implies that the shrinking of the binary could result in a
depletion of the DM minispike. However, a DM minispike could also be
replenished due to the self-interaction of DM particles or the adiabatic growth
of the central IMBH\citep{Peebles},\citep{Young},\citep{Inpser},\citep{I2}.
Therefore, we conclude with caution that the merger rates
calculated earlier in this paper are valid if DM minispike have steep profiles
$\alpha\ga2$.

\begin{figure}[!h] \centering
\includegraphics[height=4.0in]{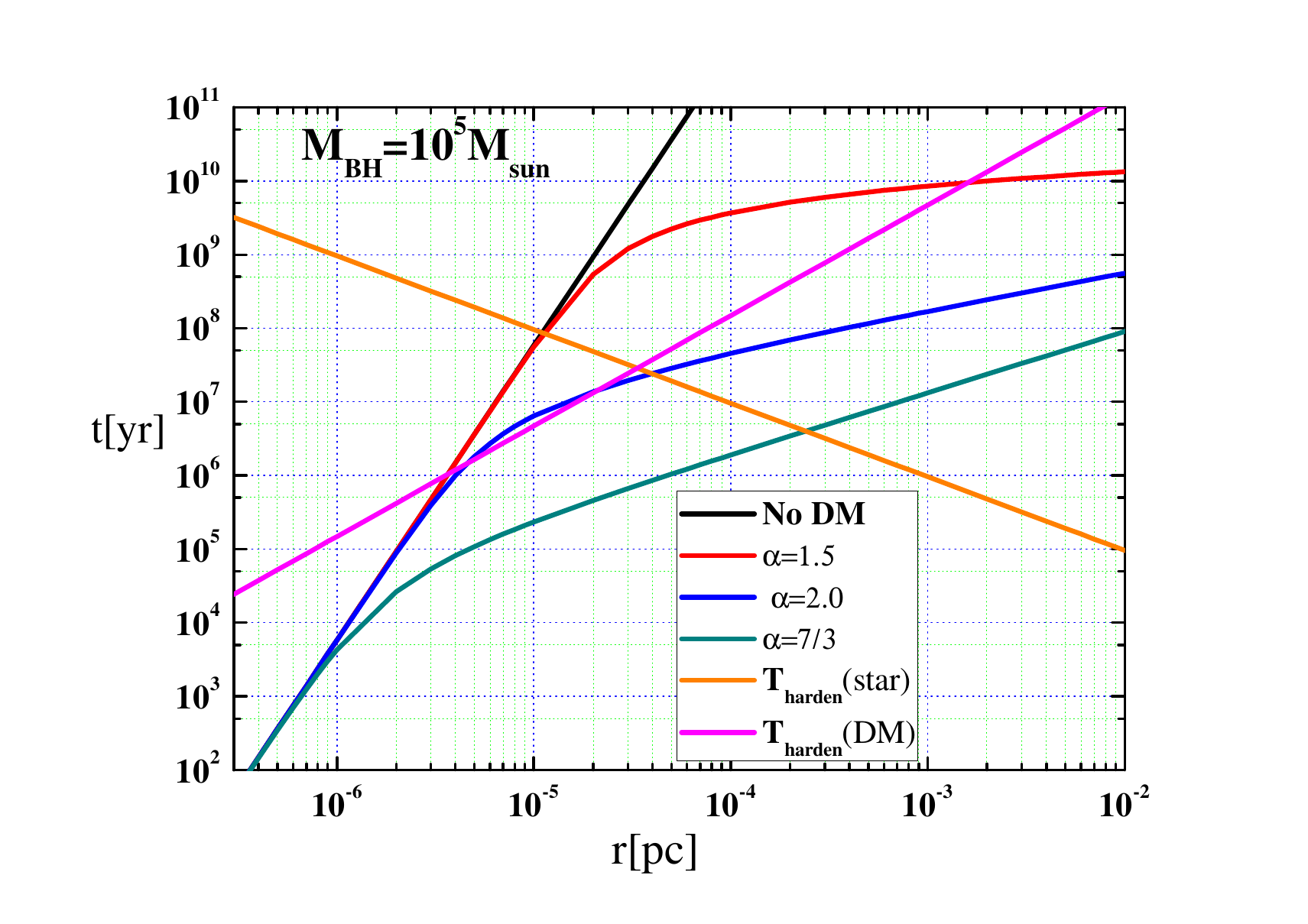} \caption{
The same as Figure~\ref{fig03} but also showing the hardening timescale
due to ejecting DM particles.
}
\label{fig10} \end{figure}

Another stellar dynamical effect that might potentially deplete the DM
minispike is called "mass segregation". Such an effect normally leads to the
formation of a dense cusp, composed of the heaviest stars in the cluster, around
an IMBH  \citep{Bahcall}.  If the cusp is mainly stellar BHs (single
population), the density would follow a power-law distribution $\rho\propto
r^{-7/4}$ (Bachall-Wolf cusp). In the conventional models of star clusters,
such a cusp forms at the expense of repelling other less massive stars, as well
as DM particles since they are also light in mass.

We think the effect of mass segregation is not important for
most of our models because the region of our interest is normally very close to
the central IMBH, so that very few stellar BHs would reach there.  More
precisely, according to Figures~\ref{fig01}-\ref{fig03}, the region where DM
plays a crucial role in triggering the mergers is where we have $T<T_{\rm
harden}$. If we define the critical radius where $T=T_{\rm harden}$ as $r_{\rm
cri}$, and calculate the number of stellar MBH inside it, we find that the
number is relatively small. For example, even if we consider a dense
Bachall-Wolf cusp, the number of BHs inside $r_{\rm cri}$ would be
\begin{equation}
N(<r_{\rm cri})=\frac{1}{10M_\odot}\int_0^{r_{\rm cri}} n_*m_\ast(r/r_{\rm infl})^{-7/4}dr,\label{015}
\end{equation}
where $r_{\rm infl}=GM/\sigma^2$ is the influence radius of the central IMBH,
and $n_*$ is the number density of stars as in Section~\ref{s4}. Then we find that given the $M_{\rm BH}$ of our interest,
the number of stellar BHs inside $r_{\rm cri}$ is $\leq1$ when $\alpha=1.5$.
When $\alpha=2.0$, this number becomes $\sim 0.5$ for $M=1000M_\odot$, $\sim 1$
for $M=10^4M_\odot$, and $\sim 6$ for $M=10^5M_\odot$.When $\alpha=7/3$, the number is $\sim 2$ for
$M=1000M_\odot$, $\sim 15$ for $M=10^4M_\odot$, and $\sim 48$ for
$M=10^5M_\odot$.

Therefore, it seems that only when $\alpha=7/3$ dose the effect
of mass segregation become relevant.  However, even in this case, if we further
compare the timescale for the Bachall-Wolf cusp to regrow and the timescale for
the stellar BHs to merger with the central IMBH ($T$ in the previous sections),
we find that the latter timescale is typically shorter.  For example, according
to Equation~(\ref{14}), the growth timescale for the cusp is $3\times 10^5\rm
yr$ for $M=1000M_\odot$, $2\times 10^6\rm yr$ for $M=10^4M_\odot$ and
$1.5\times 10^7\rm yr$ for $M=10^5M_\odot$.  They are shorter than the merger
timescales as shown in Figures~\ref{fig01}, \ref{fig02}, and \ref{fig03}.  This
result indicates that the Bachall-Wolf cusp cannot form in the case of
$\alpha=7/3$, so that we do not need to consider the effect of mass segregation
on the depletion of DM minispikes.  

\section{Conclusions}\label{s7}

In this paper we study the effect of DM minispikes around IMBHs on the merger rate
of IMRIs and EMRIs. We considered IMBHs with a mass between $10^3M_\odot$ and $10^5M_\odot$, as well as three typical density profiles for the DM, namely
$\alpha=1.5$, $2.0$, and $7/3$.

\comment{ We find that the DM minispike can decrease the merger time
dramatically, which lead to an increment of the merger rate of IMRIs in  GCs
and NSCs.}

We find that the presence of DM minispikes significantly reduces the
merger timescale of EMRIs and IMRIs due to the effet of dynamical friction. The
effect is more significant for small IMBHs ($<10^4M_\odot$) with steep DM
density profiles $\alpha\ge2.0$, and in the most extreme case the shortening of
the timescale can be two to three orders of magnitude.

As a result, the merger rate of stellar-mass BHs with IMBH in a cluster is
enhanced by as much as two orders of magnitude, to a degree that all the BHs
in the cluster are exhausted. After the BHs are depleted, the mergers with NSs
would become important and dominate the current event rate of IMRIs/EMRIs.

The enhancement of the merger rate also modifies the mass function of
IMBHs because they can grow a significant amount of mass during one Hubble
time.   This effect is important for the IMBHs with a mass of $\la10^4M_\odot$
and we derived the final mass function in this mass range according to our
model.

We conclude that the presence of DM around IMBHs would significantly change
the event rate of IMRIs and EMRIs, The effect is not necessarily an enhancement
of the merger rate because stellar-mass BHs could be exhausted and the
subsequently IMRI/EMRI events are dominated by the mergers with NSs.  On the
other hand, in the absence of DM minispikes, we predict that there are almost no
mergers of IMBHs with NSs.

The above predictions can be tested in the future by space-based GW detectors,
such as LISA, Taiji, and Tianqin. These future observations will allow us to better
understand the DM physics, the formation and evolution of IMBHs, as well as
and stellar dynamical processes in GCs and NSCs.

\section*{Acknowledgements}{This work is supported by NSFC grant Nos.11773059,
11690023,and 11873022, and by the Key Research Program of Frontier Sciences, CAS,
No.  QYZDB-SSW-SYS016. This work made use of the High Performance Computing
Resource in the Core Facility for Advanced Research Computing at Shanghai
Astronomical Observatory.}




\end{document}